\documentclass[review]{elsarticle}
\usepackage{latexsym}
\usepackage[dvipdfm,colorlinks=true,linkcolor=cyan,anchorcolor=cyan,citecolor=cyan]{hyperref}
\usepackage[hmargin={2.54cm,2.54cm},vmargin={3.17cm,3.17cm}]{geometry}
\usepackage{float}
\usepackage{graphicx,amsfonts,amsmath,amssymb,amsthm,amscd}
\usepackage{epsfig}
\usepackage{url}
\usepackage{subfigure}
\usepackage{bm}
\usepackage{xr}
\usepackage{indentfirst}
\usepackage{lineno,hyperref}

\journal{Physica A}

\begin{document}

\begin{frontmatter}

\title{\bf Optimization of a class of heat engines with explicit solution
}

\author{Yunxin Zhang\fnref{myfootnote}}
\address{Laboratory of Mathematics for Nonlinear Science, Shanghai Key Laboratory for Contemporary Applied Mathematics, Centre for Computational Systems Biology, School of Mathematical Sciences, Fudan University, Shanghai 200433, China.}
\fntext[myfootnote]{Email: xyz@fudan.edu.cn}

\begin{abstract}
A specific class of stochastic heat engines driven cyclically by time-dependent potential, which is defined in the half-line ($0<x<+\infty$), is analysed. For such engines, most of their physical quantities can be obtained explicitly, including the entropy and internal energy of the heat engine, as well as output work (power) and heat exchange with the environment during a finite time interval. The optimisation method based on the external potential to reduce {\it irreversible} work and increase energy efficiency is presented. With this optimised potential, efficiency $\eta^*$ and its particular value at maximum power $\eta^*_{\rm EMP}$ are calculated and discussed briefly.
\end{abstract}


\end{frontmatter}

\section{Introduction}
One of the hot topics in stochastic thermodynamics is the study of stochastic heat engines \cite{Sekimoto2010,Seifert2012Stochastic,Mart2016Colloidal,Giuliano2017}. According to Carnot, for heat engines working between two heat baths at temperatures $T_h>T_c$, the second law gives an upper bound $\eta_c=1-T_c/T_h$ for energy efficiency. However, this upper bound can only be achieved in quasistatic limit, where transitions occur infinitesimally slowly. Thus, the output power vanishes, which is meaningless in practice. In recent decades, many studies have discussed the efficiency at maximum power (EMP) $\eta_{\rm EMP}$.
As an important milestone, the Curzon-Ahlborn efficiency at maximum power $\eta_{CA}=1-\sqrt{T_c/T_h}$ is derived \cite{Yvon1955,Chambadal1957,Novikov1958,Curzon1975,Broeck2005}.
For the low-dissipation Carnot engine, where the entropy production per work cycle is assumed to be inversely proportional to isothermal durations, the bound for EMP $\eta_c/2\le\eta_{\rm EMP}\le \eta_c/(2-\eta_c)$ is obtained in \cite{Esposito2010Efficiency,Izumida2011Efficiency}. Meanwhile, on the basis of an analytically solvable model and scaling skills, formulation $\eta_{\rm EMP}=\eta_c/(2-\alpha\eta_c)$, with $0<\alpha<1$, is obtained in \cite{Schmiedl2008Efficiency1}. The upper bound for efficiency at arbitrary output power is discussed in \cite{Holubec2016,Ryabov2016}. General expressions for maximum power and maximum efficiency are obtained via methods of linear irreversible thermodynamics \cite{Proesmans2016}. Finally, in several recent studies, methods to operate heat engines infinitely close to the Carnot bound but at nonzero power are also suggested, either theoretically or experimentally \cite{Hondou2000Unattainability,Benenti2011,Allahverdyan2013,Verley2014The,Mart2016Brownian,Lee2017Carnot,Johnson2017Approaching,Polettini2017,Pietzonka2018,Holubec2018}.

One of the main difficulties in the study of stochastic heat engines is that, except for few specific cases \cite{Schmiedl2008Efficiency1,Schmiedl2007,Holubec2014,Holubec2015}, no explicit expressions for output work $W$, power $P$ and efficiency $\eta$ can be obtained. Therefore, numerical calculations are usually employed to find detailed properties, because no  physical quantities can be obtained explicitly, or the corresponding expressions are too complicated to be used to derive more meaningful results theoretically \cite{Then2008,Horowitz2018}.

One possible reason that the energy efficiency $\eta$ of heat engines at
nontrivial power is usually less than the Carnot efficiency $\eta_c$ is that  nonzero {\it irreversible} work is usually spent during the work cycle of heat engines to overcome the viscous friction in the environment: the more the {\it irreversible} work spent, the lower the energy efficiency \cite{Giuliano2017, Schmiedl2008Efficiency,Zhang20091}. The optimisation of heat engines to reduce {\it irreversible} work as much as possible is one of the main aims of this study. In general, this goal is difficult to accomplish analytically.  In this study, a specific class of stochastic heat engines is presented, for which most of the physically interesting quantities can be obtained in simple form. With these simple and explicit expressions, heat engines can be optimised to attain their highest output work and highest energy efficiency.

For a stochastic heat engine driven by time-dependent potential $V(x,\tau)$, the probability density $p(x, \tau)$ to find it in state (internal degree of freedom) $x$ at time $\tau$ is governed by the following Fokker-Planck equation, see \cite{Schmiedl2008Efficiency},
\begin{eqnarray}\label{eqFP}
\partial_\tau p(x,\tau)=\partial_x[p(x,\tau)\partial_x V(x,\tau)/\gamma+D\partial_x p(x,\tau)].
\end{eqnarray}
Where $\gamma$ is the drag coefficient, and $D$ is the free diffusion constant, which satisfies $\gamma D=k_BT$, with $k_B$ as the Boltzmann constant and $T$ as the absolute temperature. Both $\gamma$ and $D$ might be time $\tau$ dependent.

\begin{figure}
  \includegraphics[width=15cm]{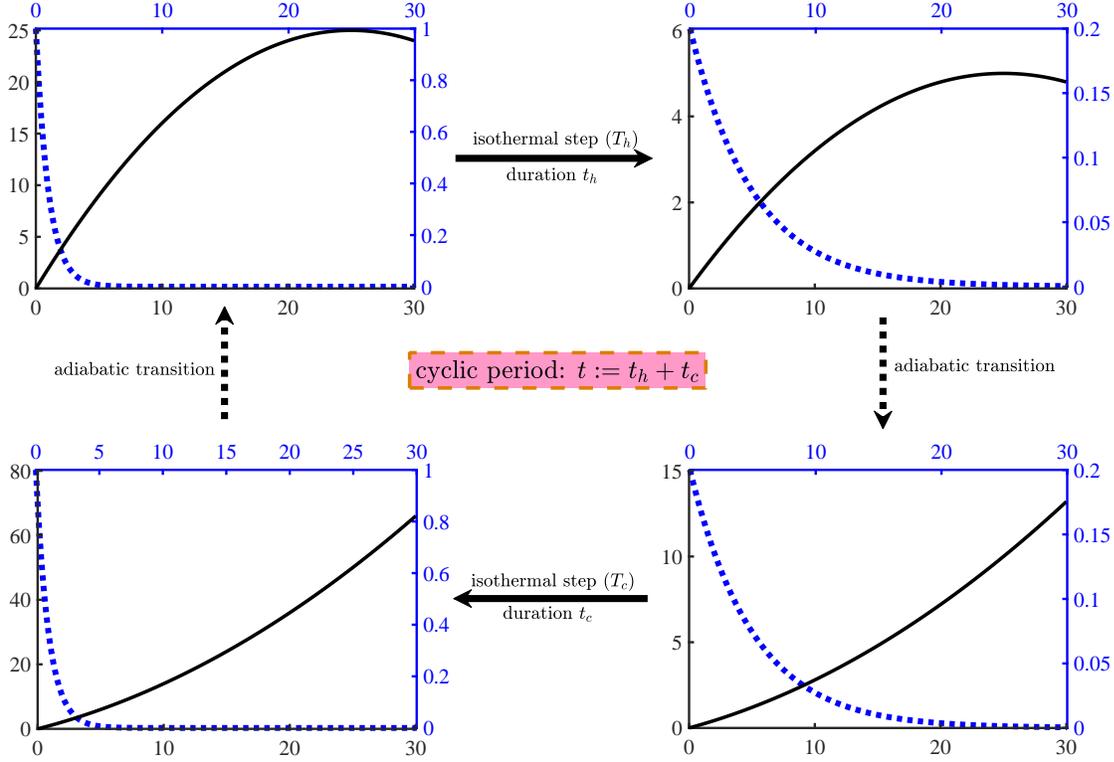}\\
  \caption{(Color online) Schematic depiction of a stochastic heat engine based on a particle in a time-dependent potential $V(x,\tau)$. The potential $V(x,\tau)$ is plotted by a solid line (with left and bottom labels), and the probability density $p(x,\tau)$ is plotted by dotted line (with right and top labels). Parameter values used in calculations are $t_h=t_c=1, \gamma(0)=\gamma(t_h)=\gamma(t_h+t_c)=0.02, \lambda(0)=\lambda(t_h+t_c)=1, \lambda(t_h)=5, \lambda'(\tau)=[\lambda(t_h)-\lambda(0)]/t_h$ for $0<\tau<t_h$, and $\lambda'(\tau)=-[\lambda(t_h)-\lambda(0)]/t_c$ for $t_h<\tau<t_h+t_c$. Temperatures in the two isothermal steps are $T_h=2T_c=2$. See Eqs.~(\ref{potential1},\ref{density1}) for expressions for potential $V(x,\tau)$ and probability density $p(x,\tau)$ used in calculations. The four subfigures correspond to time $\tau=0+, t_h-, t_h+$, and $t-$. During the isothermal step with temperature $T_h$, parameter $\lambda$ increases with time $\tau$. Therefore, the potential $V(x,\tau)$ becomes flatter with time, and the probability density $p(x,\tau)$ becomes wider, which then leads to the output of work. On the contrary, during the isothermal step with temperature $T_c$, parameter $\lambda$ decreases with time $\tau$ and then leads to the input of work. }\label{shematic}
\end{figure}

Similar to previous studies \cite{Schmiedl2008Efficiency,Holubec2014},  stochastic heat engines discussed in this study are assumed to work cyclically, with two isothermal processes and two adiabatic transitions. Fig.~\ref{shematic} shows a sketch of an operational cycle. In each work cycle, the engines perform sequentially through the following four steps \cite{Schmiedl2008Efficiency}: {\bf (1)} isothermal process with high temperature $T_h$ during time interval $0<\tau<t_h$, {\bf (2)} adiabatic transition (instantaneously) from high temperature $T_h$ to low temperature $T_c$ at time $\tau=t_h$, {\bf (3)} isothermal process with low temperature $T_c$ during time interval $t_h<\tau<t_h+t_c$ and,  {\bf (4)} adiabatic transition from low temperature $T_c$ to high temperature $T_h$ at time $\tau=t_h+t_c$. As in \cite{Schmiedl2008Efficiency,Holubec2014}, adiabatic transitions are idealised as sudden jumps of potential and assumed to occur instantaneously without heat exchange. The period of work cycle is denoted by $t:=t_h+t_c$.

\section{Stochastic heat engines with explicit solution}\label{case1}
For given potential
\begin{eqnarray}\label{potential1}
V(x,\tau)=-\frac{\gamma(\tau) \lambda'(\tau)}{2\lambda(\tau)}x^2+\frac{k_BT(\tau)}{\lambda(\tau)}x,
\end{eqnarray}
with $0\le x<+\infty$, we can easily verify that
\begin{eqnarray}\label{density1}
p(x,\tau)=\frac{1}{\lambda(\tau)}\exp\left(-\frac{x}{\lambda(\tau)}\right), \quad 0\le x<+\infty,
\end{eqnarray}
is a solution of Eq.~(\ref{eqFP}), where $\lambda(\tau)>0$ is a time-dependent parameter that is used to regulate the potential $V(x,\tau)$. In this study, the state (spatial) variable $x$ is defined in interval $[0, +\infty)$, with $x=0$ as a reflecting boundary such that $\int_0^{+\infty}p(x,\tau)\equiv1$ is satisfied at any time $\tau$.
Notably, if parameter $\lambda$ is time independent ({\it i.e.}, $\lambda'\equiv0$), $p(x,\tau)$ given in Eq.~(\ref{density1}) is the equilibrium (Boltzmann) probability density corresponding to the potential given only by the second term of Eq.~(\ref{potential1}). Given the reflecting boundary condition at $x=0$, the second term of Eq.~(\ref{potential1}) is not linear, so the probability density $p(x,\tau)$ given in Eq.~(\ref{density1}) is not in Gaussian type as expected.

The probability density $p(x,\tau)$ given in Eq.~(\ref{density1}) is not a general solution of Eq.~(\ref{eqFP}), it is only a time periodic solution obtained by appropriate initial condition $p(x,0)$. With an arbitrary initial condition $p(x,0)$, the solution of Eq.~(\ref{eqFP}) may or may not converge at a long time to the one given in Eq.~(\ref{density1}). Although $p(x,\tau)$ given in Eq.~(\ref{density1}) is temperature independent, the thermodynamic properties of the corresponding machine are sensitive to the bath temperature; hence, it can be regarded as a heat engine. The influence of bath temperature enters through potential $V(x,\tau)$, which is different from other heat engines as discussed previously in references.

From Eq.~(\ref{potential1}), or Fig.~\ref{shematic}, one may find that the potential $V(x,\tau)$ decreases to $-\infty$ for large $x$ and $\lambda'>0$. Thus, the system may escape to $+\infty$. However, this phenomenon occurs only during parts of the work cycle when parameter $\lambda$ increases, and the system does not have sufficient time to escape to $+\infty$. During another isothermal step, $\lambda'$ will change from positive to negative, and potential $V(x,\tau)$ will increase to $+\infty$ for large $x$, so the system will accumulate toward 0 again.

By definition, the system entropy at time $\tau$ is
\begin{eqnarray}\label{entropy1}
S(\tau)=-k_B\int_0^{+\infty}p(x,\tau)\ln p(x,\tau)dx
=k_B[\ln\lambda(\tau)+1].
\end{eqnarray}
The mean internal energy at time $\tau$ is
\begin{eqnarray}\label{energy1}
E(\tau)=\int_0^{+\infty}p(x,\tau)V(x,\tau)dx
=k_BT(\tau)-\gamma(\tau)\lambda(\tau)\lambda'(\tau).
\end{eqnarray}
In particular for $\lambda'(\tau)\equiv0$, the probability density $p(x,\tau)$ is the steady state (equilibrium) solution of Eq.~(\ref{eqFP}). Therefore, internal energy $E(\tau)$ is proportional to the bath temperature $T$.

The mean work extracted during time interval $[t_i, t_f]$ is
\begin{eqnarray}\label{work1}
W(t_i, t_f)&=&-\int_{t_i}^{t_f}\int_0^{+\infty}p(x,\tau)\partial_\tau V(x,\tau)dxd\tau \nonumber\\
&=&E(t_i)-E(t_f)+\int_{t_i}^{t_f}T(\tau)dS(\tau)
-2\int_{t_i}^{t_f}\gamma(\tau)(\lambda'(\tau))^2d\tau.
\end{eqnarray}
Therefore, the heat uptake from heat bath during time interval $[t_i, t_f]$ is
\begin{eqnarray}\label{heat1}
Q(t_i, t_f)&=&\int_{t_i}^{t_f}\int_0^{+\infty}\partial_\tau p(x,\tau)V(x,\tau)dxd\tau
=W(t_i, t_f)+E(t_f)-E(t_i)\nonumber\\
&=&\int_{t_i}^{t_f}T(\tau)dS(\tau)-2\int_{t_i}^{t_f}\gamma(\tau)(\lambda'(\tau))^2d\tau.
\end{eqnarray}
Here, for simplicity, we assume that integral $\int_0^{+\infty}\partial_\tau p(x,\tau)V(x,\tau)dx>0$ for any $\tau\in[t_i,t_f]$. Otherwise, $Q(t_i, t_f)$ is less than the value of heat uptake, but the value of heat uptake can be obtained by the same methods as in \cite{Holubec2014}.

Similar to \cite{Schmiedl2008Efficiency1}, the mean {\it irreversible} work spent by the heat engine during time interval $[t_i, t_f]$ can be calculated by $W^{\rm irr}(t_i, t_f)=\int_{t_i}^{t_f}\int_0^{+\infty}\gamma(\tau)j^2(x,\tau)/p(x,\tau)dxd\tau
=\int_{t_i}^{t_f}\int_0^{+\infty}\gamma(\tau)p(x,\tau)v^2(x,\tau)dxd\tau$, where $j(x,\tau)=p(x,\tau)v(x,\tau)$ is the flux of probability at time $\tau$, and  $v(x,\tau)=-\partial_x[V(x,\tau)+k_BT(\tau)\ln p(x,\tau)]/\gamma(\tau)$ is the instantaneous speed. For this specific case with potential $V(x,t)$ as given in Eq.~(\ref{potential1}),
\begin{eqnarray}\label{irreversible1}
W^{\rm irr}(t_i, t_f)
=2\int_{t_i}^{t_f}\gamma(\tau)(\lambda'(\tau))^2d\tau
=\int_{t_i}^{t_f}T(\tau)dS(\tau)-Q(t_i, t_f).
\end{eqnarray}
Or, $Q(t_i, t_f)=\int_{t_i}^{t_f}T(\tau)dS(\tau)-W^{\rm irr}(t_i, t_f)$.

For convenience, denote $G_h=G(0, t_h)$ and $G_c=G(t_h, t_h+t_c)$ for $G=W, Q, W^{\rm irr}$. Here, $W_{h/c}$, $Q_{h/c}$ and $W^{\rm irr}_{h/c}$ are the output work, heat uptake from heat bath and {\it irreversible} work spent during the isothermal process with temperature $T_{h/c}$, respectively. The total output work per work cycle of heat engine with period $t=t_h+t_c$ is
\begin{eqnarray}\label{worktot1}
W^{\rm tot}&:=&W(0, t)=W_h+W_c
 =\int_{0}^{t}T(\tau)dS(\tau)-2\int_{0}^{t}\gamma(\tau)(\lambda'(\tau))^2d\tau =Q_h+Q_c.
\end{eqnarray}

Eq.~(\ref{irreversible1}) shows that the {\it irreversible} work $W^{\rm irr}(t_i, t_f)$ is a functional of $\lambda(\tau)$, which is the parameter used to regulate external potential $V(x,\tau)$ (Eq.~(\ref{potential1})).
The variation of $W^{\rm irr}(t_i, t_f)$ according to $\lambda(\tau)$ is
\begin{eqnarray}\label{variation1}
\delta W^{\rm irr}(t_i, t_f)&=&4\int_{t_i}^{t_f}[\gamma(\tau)\lambda'(\tau)]'\delta \lambda(\tau) d\tau,
\end{eqnarray}
where $\delta\lambda(\tau)$ is an arbitrary variation of parameter (function) $\lambda(\tau)$, which satisfy $\delta\lambda(t_i)=\delta\lambda(t_f)=0$. Here, in variational process, $\lambda_i:=\lambda(t_i), \lambda_f:=\lambda(t_f)$ are assumed to maintain {\it initial} entropy $S(t_i)$ and the {\it final} entropy $S(t_f)$ are unchanged (Eq.~(\ref{entropy1})). Thus, if the temperature $T$ is time independent during time interval $[t_i, t_f]$, then the value of $\int_{t_i}^{t_f}TdS(\tau)=T[S(t_f)-S(t_i)]=:T\Delta S(t_i, t_f)$ is not influenced by the function variation $\delta\lambda(\tau)$. Therefore, from Eq.~(\ref{heat1}), the variation of heat uptake $Q(t_i, t_f)$ is $\delta Q(t_i, t_f)=-\delta W^{\rm irr}(t_i, t_f)$. Consequently, for given values of $\lambda_0:=\lambda(0)\equiv\lambda(t)$ and $\lambda_1:=\lambda(t_h)$, the maximal value of total work $W^{\rm tot}$ is reached if and only if the {\it irreversible} work $W^{\rm irr}(0, t)=W^{\rm irr}_h+W^{\rm irr}_c$ reaches its minimal value. Notably, $\delta W^{\rm tot}=\delta Q_h+\delta Q_c=-\delta W^{\rm irr}_h-\delta W^{\rm irr}_c=-\delta W^{\rm irr}(0,t)$, see Eq.~(\ref{worktot1}).

With the optimal function $\lambda^*(\tau)$, variation $\delta W^{\rm irr}(t_i, t_f)\equiv0$ for any $\delta\lambda^*(\tau)$. Thus $[\gamma(\tau)\lambda^{*\prime}(\tau)]'=0$ should be satisfied (Eq.~(\ref{variation1})). The optimal parameter $\lambda^*(\tau)$ can be obtained as follows,
\begin{eqnarray}\label{optimalpara1}
\lambda^*(\tau)=\frac{\lambda_i\int_{\tau}^{t_f}\frac{d\hat\tau}{\gamma(\hat\tau)}
+\lambda_f\int_{t_i}^{\tau}\frac{d\hat\tau}{\gamma(\hat\tau)}}{\int_{t_i}^{t_f}\frac{d\hat\tau}{\gamma(\hat\tau)}},
\quad t_i\le\tau\le t_f.
\end{eqnarray}
In particular, for constant drag coefficient $\gamma(\tau)\equiv\gamma_0$, the optimal parameter
$\lambda^*(\tau)=[\lambda_i(t_f-\tau)+\lambda_f(\tau-t_i)]/(t_f-t_i)$, which changes linearly with time $\tau$. With $\lambda^*(\tau)$, the optimal potential $V^*(x,\tau)$ can be obtained by Eq.~(\ref{potential1}), and the corresponding probability density $p^*(x,\tau)$ can be obtained by Eq.~(\ref{density1}).

As shown in Eqs.~(\ref{irreversible1}) and (\ref{optimalpara1}), the minimal value of {\it irreversible} work spent during time interval $[t_i, t_f]$ can be obtained as follows,
\begin{eqnarray}\label{irreversiblemin1}
W^{\rm irr*}(t_i, t_f)
=\frac{2(\lambda_f-\lambda_i)^2}{\int_{t_i}^{t_f}\frac{d\tau}{\gamma(\tau)}}.
\end{eqnarray}
For the special case that $\gamma(\tau)\equiv\gamma_0$ is constant, $W^{\rm irr*}(t_i, t_f)=2\gamma_0(\lambda_f-\lambda_i)^2/(t_f-t_i)$. Which gives that
\begin{eqnarray}\label{irreversibleminhc1}
W^{\rm irr*}_h=2\gamma_h(\lambda_1-\lambda_0)^2/t_h,\quad
W^{\rm irr*}_c=2\gamma_c(\lambda_0-\lambda_1)^2/t_c.
\end{eqnarray}
In this equation, $\gamma_h$ and $\gamma_c$ are drag coefficients corresponding to the isothermal process with high and low temperatures, respectively.

In summary, with constant temperatures $T_h$ and $T_c$, constant drag coefficients $\gamma_h$ and $\gamma_c$ and the optimal parameter $\lambda^*(\tau)$, the output work, heat uptake from hot heat bath and efficiency can be obtained as follows (see Eqs.~(\ref{heat1}) and (\ref{worktot1})),
\begin{eqnarray}\label{efficiency1}
&&W^{\rm tot*}=(T_h-T_c)\Delta S-2(\lambda_1-\lambda_0)^2\left(\frac{\gamma_h}{t_h}+
\frac{\gamma_c}{t_c}\right), \nonumber \\
&&Q_h^*=T_h\Delta S-\frac{2(\lambda_1-\lambda_0)^2\gamma_h}{t_h}, \nonumber \\
&&\eta^*:=\frac{W^{\rm tot*}}{Q_h^*}
=1-\frac{T_c\Delta S+\frac{2(\lambda_1-\lambda_0)^2\gamma_c}{t_c}}{T_h\Delta S-\frac{2(\lambda_1-\lambda_0)^2\gamma_h}{t_h}}\nonumber \\
&&\ \ \quad=1-\frac{T_c}{T_h}\cdot
\frac{1+\Omega \gamma_c/T_ct_c}{1-\Omega \gamma_h/T_ht_h}.
\end{eqnarray}
Where $\Delta S=\Delta S(0,t_h)=S(t_h)-S(0)=k_B\ln(\lambda_1/\lambda_0)$, and $\Omega=2(\lambda_1-\lambda_0)^2/\Delta S$. One can easily verify that the {\it stall} time $t_{stall}$ with which the output work is vanished, is
\begin{eqnarray}\label{stalltime1}
t_{stall}=\frac{2(\lambda_1-\lambda_0)^2}{(T_h-T_c)\Delta S}\left(\frac{\gamma_h}{\bar t_h}+
\frac{\gamma_c}{\bar t_c}\right),
\end{eqnarray}
where $\bar t_{h/c}=t_{h/c}/t$, with $t=t_h+t_c$ the cyclic period of engine. Note that, $\bar t_{h/c}$ depends only on the ratio $t_h/t_c$, and it is independent of cyclic period $t$. The power $P^*=W^{\rm tot*}/t$ reaches its maximal value $P^*_{\max}$ when cyclic period $t=t^*:=2t_{stall}$, and
$P^*_{\max}=(T_h-T_c)\Delta S/(4t_{stall})$. The efficiency at maximum power (EMP) is then
\begin{eqnarray}\label{EMP1}
\eta^*_{\rm EMP}=\frac{W^{\rm tot*}(t=2t_{stall})}{Q_h^*(t=2t_{stall})}=\frac{\eta_c}{2-\frac{\eta_c\gamma_h/\bar t_h}{\gamma_h/\bar t_h+\gamma_c/\bar t_c}}=:\frac{\eta_c}{2-\alpha\eta_c},
\end{eqnarray}
with $\eta_c=1-T_c/T_h$ the Carnot efficiency and $\alpha={\gamma_h\bar t_c}/{(\gamma_h\bar t_c+\gamma_c\bar t_h)}$. Obviously, $\eta_c/2\le \eta^*_{\rm EMP}\le \eta_c/(2-\eta_c)$, which is the same as obtained previously in \cite{Schmiedl2008Efficiency,Esposito2010Efficiency,Izumida2011Efficiency}.
Note, $\bar t_h+\bar t_c=1$ and $0<\alpha<1$.

\section{More general cases}\label{caseGeneral}
In general for given potential{\small
\begin{eqnarray}\label{potentialGeneral}
V(x,\tau)=\frac{k_BT(\tau)}{n\sigma(\tau)}x^n-\frac{\gamma(\tau)\sigma'(\tau)}{2n\sigma(\tau)}x^2-(n-1)k_BT(\tau)\ln x,\qquad 0\le x<+\infty,
\end{eqnarray}}
we can easily verified that one solution of Eq.~(\ref{eqFP}) is
\begin{eqnarray}\label{densityGeneral}
p(x,\tau)=\frac{x^{n-1}}{\sigma(\tau)}\exp\left(-\frac{x^n}{n\sigma(\tau)}\right), \qquad 0\le x<+\infty,
\end{eqnarray}
where $n\ge1$ is an integer number. $\sigma(\tau)$ is a parameter (function) used to regulate potential $V(x,\tau)$.

Similar to the special cases discussed in previous section, if $\sigma'(\tau)\equiv0$ then $p(x,\tau)$ given by Eq.~(\ref{densityGeneral}) is the equilibrium solution of Eq.~(\ref{eqFP}) but with potential $V(x,\tau)={k_BT(\tau)}x^n/{n\sigma(\tau)}-(n-1)k_BT(\tau)\ln x$. For $\sigma'(\tau)>0$, potential $V(x,\tau)$ decreases to $-\infty$ for large $x$, so one may think the system will escape to $+\infty$. However this occurs only during parts of the work cycle. During other parts of the cycle, $\sigma'(\tau)$ will change from positive to negative, therefore, the potential $V(x,\tau)$ increases to $+\infty$ for large $x$. The system will move toward 0 again. Meanwhile, the influence of bath temperature enters through the potential $V(x,\tau)$ but not through probability density $p(x,\tau)$ directly. The machine driving by potential $V(x,\tau)$ given in Eq.~(\ref{potentialGeneral}) can also be regarded as a heat engine.

By definition, the system entropy at time $\tau$ is \begin{eqnarray}\label{entropyGeneral}
S(\tau)&=&-k_B\int_0^{+\infty}p(x,\tau)\ln p(x,\tau)dx \nonumber\\
&=&k_B\left[\frac1n\ln\sigma(\tau)+1-\frac{n-1}{n}\ln{n}\right.
\left.-\frac{n-1}{n}\int_0^{+\infty}e^{-x}\ln xdx\right].
\end{eqnarray}
The mean internal energy at time $\tau$ is {\small
\begin{eqnarray}\label{energyGeneral}
E(\tau)&=&\int_0^{+\infty}p(x,\tau)V(x,\tau)dx \nonumber\\
&=&k_BT(\tau)-\Gamma\left(\frac2n+1\right)\frac{\gamma(\tau)\sigma'(\tau)}{2}[n\sigma(\tau)]^{2/n-1}
-\frac{(n-1)k_BT(\tau)}{n}\ln[{n\sigma(\tau)}] \nonumber\\
&&-\frac{(n-1)k_BT(\tau)}{n}\int_0^{+\infty}e^{-x}\ln xdx,
\end{eqnarray}}
where $\Gamma(s):=\int_0^{+\infty}x^{s-1}\exp(-x)dx$ is the {\it Gamma} function.
The mean work extracted during time interval $[t_i, t_f]$ (see Eqs.~(\ref{potentialGeneral}) and (\ref{densityGeneral})){\footnotesize
\begin{eqnarray}\label{workGeneral}
W(t_i, t_f)&=&-\int_{t_i}^{t_f}\int_0^{+\infty}p(x,\tau)\partial_\tau V(x,\tau)dxd\tau \nonumber\\
&=&E(t_i)-E(t_f)+\int_{t_i}^{t_f}k_BT(\tau)dS(\tau)
-\Gamma\left(\frac2n+1\right)\int_{t_i}^{t_f}[n\sigma(\tau)]^{\frac2n-2}\gamma(\tau)[\sigma'(\tau)]^2d\tau.
\end{eqnarray}}
The heat uptake from heat bath during time interval $[t_i, t_f]$ is {\footnotesize
\begin{eqnarray}\label{heatGeneral}
Q(t_i, t_f)&=&W(t_i, t_f)+E(t_f)-E(t_i)\nonumber\\
&=&\int_{t_i}^{t_f}k_BT(\tau)dS(\tau)
-\Gamma\left(\frac2n+1\right)\int_{t_i}^{t_f}[n\sigma(\tau)]^{2/n-2}\gamma(\tau)[\sigma'(\tau)]^2d\tau.
\end{eqnarray}}
For such cases, the {\it irreversible} work spent in time interval $[t_i, t_f]$ is {\small
\begin{eqnarray}\label{irreversibleGeneral}
W^{\rm irr}(t_i, t_f)&=&\int_{t_i}^{t_f}\int_0^{+\infty}\gamma(\tau)j^2(x,\tau)/p(x,\tau)dxd\tau \nonumber \\
&=&\Gamma\left(\frac2n+1\right)\int_{t_i}^{t_f}[n\sigma(\tau)]^{2/n-2}\gamma(\tau)[\sigma'(\tau)]^2d\tau \nonumber \\
&=&\int_{t_i}^{t_f}T(\tau)dS(\tau)-Q(t_i, t_f).
\end{eqnarray}}
The total output work per work cycle with period $t=t_h+t_c$ is (see Eq.~(\ref{workGeneral}))
\begin{eqnarray}\label{worktotGeneral}
W^{\rm tot}:=W(0, t)
=\int_{0}^{t}k_BT(\tau)dS(\tau)-W^{\rm irr}(0, t).
\end{eqnarray}

The variation of {\it irreversible} work $W^{\rm irr}(t_i, t_f)$ according to parameter $\sigma(\tau)$ is
{\small
\begin{eqnarray}
\label{variationGeneral}
\delta W^{\rm irr}(t_i, t_f)&=&-\Gamma\left(l+3\right)n^l\int_{t_i}^{t_f}
\left\{l\gamma(\tau)[\sigma(\tau)]^{l-1}[\sigma'(\tau)]^2\right.\nonumber\\
&&+2\gamma'(\tau)[\sigma(\tau)]^{l}\sigma'(\tau)
+\left.2\gamma(\tau)[\sigma(\tau)]^{l}\sigma''(\tau)
\right\}
\delta \sigma(\tau) d\tau,
\end{eqnarray}}
where $l:=2/n-2$, and $\delta\sigma(\tau)$ is an arbitrary variation of parameter (function) $\sigma(\tau)$, which satisfy $\delta\sigma(t_i)=\delta\sigma(t_f)=0$.

Similar to the above discussion about the special cases,
for given parameter values of $\sigma_0:=\sigma(0)\equiv\sigma(t)$ and $\sigma_1:=\sigma(t_h)$, the maximal value of total output work $W^{\rm tot}$ is reached, which is equivalent to the scenario that the minimal value of {\it irreversible} work $W^{\rm irr}(0, t)=W^{\rm irr}_h+W^{\rm irr}_c$ is reached, if and only if parameter (function) $\sigma(\tau)$ is equal to its optimal value $\sigma^*(\tau)$, with $\sigma^*(\tau)$ satisfying $\delta W^{\rm irr}_h=\delta W^{\rm irr}_c=0$ for any increment $\delta \sigma^*(\tau)$. From Eq.~(\ref{variationGeneral}),
$\sigma^*(\tau)$ satisfies
\begin{eqnarray*}
&&l\gamma(\tau)[\sigma^*(\tau)]^{l-1}[\sigma^{*\prime}(\tau)]^2+2\gamma'(\tau)[\sigma^*(\tau)]^{l}\sigma^{*\prime}(\tau) +2\gamma(\tau)[\sigma^*(\tau)]^{l}\sigma^{*\prime\prime}(\tau)=0\\
&&\Longleftrightarrow l\frac{\sigma^{*\prime}(\tau)}{\sigma(\tau)}+2\frac{\gamma'(\tau)}{\gamma(\tau)}+2\frac{\sigma^{*\prime\prime}(\tau)}{\sigma^{*\prime}(\tau)}=0\\
&&\Longleftrightarrow {\bf (}\ln[\gamma(\tau)\sigma^{*\prime}(\tau)(\sigma^*(\tau))^{l/2}]{\bf )}'=0\\
&&\Longleftrightarrow\left(\frac{(\sigma^*(\tau))^{l/2+1}}{l/2+1}\right)'=\frac{c_0}{\gamma(\tau)}\\
&&\stackrel{\mbox{{\tiny $l\!=\!2\!/\!n\!-\!2$}}}{\Longleftrightarrow}
\left(n(\sigma^*(\tau))^{1/n}\right)'=\frac{c_0}{\gamma(\tau)}
\\
&&\Longleftrightarrow (\sigma^*(\tau))^{1/n}=c_1\int_{t_i}^{t_f}\frac{d\hat\tau}{\gamma(\hat\tau)}+c_2.
\end{eqnarray*}
Where $c_0, c_1, c_2$ are constants determined by boundary conditions $\sigma^*(t_i)=\sigma_i$ and $\sigma^*(t_f)=\sigma_f$. By routine mathematical analysis, one can show that
\begin{eqnarray}\label{optimalparaGeneral}
\sigma^*(\tau)=\left(\frac{\sqrt[n]{\sigma_i}\int_{\tau}^{t_f}\frac{d\hat\tau}{\gamma(\hat\tau)}
+\sqrt[n]{\sigma_f}\int_{t_i}^{\tau}\frac{d\hat\tau}{\gamma(\hat\tau)}}{\int_{t_i}^{t_f}\frac{d\hat\tau}{\gamma(\hat\tau)}}\right)^n,
\end{eqnarray}
for $t_i\le\tau\le t_f$.

With the optimal parameter $\sigma^*(\tau)$, the minimal value of {\it irreversible} work spent during time interval $[t_i, t_f]$ is, see Eqs.~(\ref{irreversibleGeneral}, \ref{optimalparaGeneral})
\begin{eqnarray}\label{irreversiblemin1}
W^{\rm irr*}(t_i, t_f)
=n^{\frac{2}{n}}\Gamma\left(\frac2n+1\right)
\frac{(\sqrt[n]{\sigma_f}-\sqrt[n]{\sigma_i})^2}{\int_{t_i}^{t_f}\frac{d\tau}{\gamma(\tau)}}.
\end{eqnarray}
For special cases in which $\gamma(\tau)\equiv\gamma_0$ is constant, $W^{\rm irr*}(t_i, t_f)=n^{2/n}\Gamma(2/n+1)\gamma_0(\sqrt[n]{\sigma_f}-\sqrt[n]{\sigma_i})^2/(t_f-t_i)$, which yields
\begin{eqnarray}\label{irreversibleminhc1}
W^{\rm irr*}_h&=&n^{2/n}\Gamma(2/n+1)\gamma_h(\sqrt[n]{\sigma_1}-\sqrt[n]{\sigma_0})^2/t_h,\nonumber\\
W^{\rm irr*}_c&=&n^{2/n}\Gamma(2/n+1)\gamma_c(\sqrt[n]{\sigma_1}-\sqrt[n]{\sigma_0})^2/t_c,
\end{eqnarray}
where $\gamma_h$ and $\gamma_c$ are drag coefficients corresponding to the isothermal process with high and low temperatures, respectively.

Similar results as given  in Eqs.~(\ref{efficiency1}, \ref{stalltime1}, \ref{EMP1}) can be obtained, but with $\Delta S$ given by $\Delta S=k_B\ln\sqrt[n]{\sigma_1/\sigma_0}$, $\lambda_1-\lambda_0$ replaced by $\sqrt[n]{\sigma_1}-\sqrt[n]{\sigma_0}$, and the constant 2 in Eqs.~(\ref{efficiency1}) and (\ref{stalltime1}) replaced by $n^{2/n}\Gamma(2/n+1)$. Note, $\Gamma(k+1)=k\Gamma(k)$ and $\Gamma(1)=1$.

\section{Conclusions and remarks}
In summary, a specific class of stochastic heat engines is presented in this study, for which most of the physically interesting quantities can be obtained explicitly. With these explicit expressions, detailed properties of heat engines can be obtained.
Meanwhile, on the basis on these explicit expressions, heat engines can be optimised to achieve large output work and high energy efficiency by reducing the {\it irreversible} work spent in a work cycle.


\begin{thebibliography}{10}

\bibitem{Sekimoto2010}
Ken Sekimoto.
\newblock {\em Stochastic Energetics}.
\newblock Springer Berlin Heidelberg, 2010.

\bibitem{Seifert2012Stochastic}
U.~Seifert.
\newblock Stochastic thermodynamics, fluctuation theorems and molecular
  machines.
\newblock {\em Rep. Prog. Phys.}, 75(75):126001, 2012.

\bibitem{Mart2016Colloidal}
I.~A. Mart\'{\i}nez, \'{E}. Rold\'{a}n, L.~Dinis, and R.~A. Rica.
\newblock Colloidal heat engines: a review.
\newblock {\em Soft Matter}, 13(1):22, 2016.

\bibitem{Giuliano2017}
B.~Giuliano, G.~Casati, K.~Saito, and R.~Whitney.
\newblock Fundamental aspects of steady-state conversion of heat to work at the
  nanoscale.
\newblock {\em Phys. Rep.}, 694:1--124, 2017.

\bibitem{Yvon1955}
J.~Yvon.
\newblock {\em Proceedings of the International Conference on Peaceful Uses of
  Atomic Energy}.
\newblock United Nations, Geneva, 1955, p.~387.

\bibitem{Chambadal1957}
P.~Chambadal.
\newblock {Les Centrales Nucl\'{e}aries}.
\newblock {\em Armand Colin}, 11:41--58, 1957.

\bibitem{Novikov1958}
I.~I. Novikov.
\newblock The efficiency of atomic power stations (a review).
\newblock {\em J. Nucl. Energy}, 7:125--128, 1958.

\bibitem{Curzon1975}
F.~L. Curzon and B.~Ahlborn.
\newblock Effciency of a carnot engine at maximum power output.
\newblock {\em Phil. Mag. Ser.}, 43:22--24, 1975.

\bibitem{Broeck2005}
C.~Van den Broeck.
\newblock Thermodynamic efficiency at maximum power.
\newblock {\em Phys. Rev. Lett.}, 95(19):190602, 2005.

\bibitem{Esposito2010Efficiency}
M.~Esposito, R.~Kawai, K.~Lindenberg, and C.~Van den Broeck.
\newblock Efficiency at maximum power of low-dissipation carnot engines.
\newblock {\em Phys. Rev. Lett.}, 105(15):150603, 2010.

\bibitem{Izumida2011Efficiency}
Y.~Izumida and K.~Okuda.
\newblock Efficiency at maximum power of minimally nonlinear irreversible heat
  engines.
\newblock {\em Europhys. Lett.}, 97(1):10004, 2011.

\bibitem{Schmiedl2008Efficiency1}
T.~Schmiedl and U.~Seifert.
\newblock Efficiency at maximum power: An analytically solvable model for
  stochastic heat engines.
\newblock {\em Europhys. Lett.}, 81(2):20003, 2008.

\bibitem{Holubec2016}
V.~Holubec and A.~Ryabov.
\newblock Maximum efficiency of low-dissipation heat engines at arbitrary
  power.
\newblock {\em J. Stat. Mech.}, 2016(7):073204, 2016.

\bibitem{Ryabov2016}
A.~Ryabov and V.~Holubec.
\newblock Maximum efficiency of steady-state heat engines at arbitrary power.
\newblock {\em Phys. Rev. E}, 93(5):050101, 2016.

\bibitem{Proesmans2016}
K.~Proesmans, B.~Cleuren, and den Broeck C.~Van.
\newblock Power-efficiency-dissipation relations in linear thermodynamics.
\newblock {\em Phys. Rev. Lett.}, 116:220601, 2016.

\bibitem{Hondou2000Unattainability}
T.~Hondou and K.~Sekimoto.
\newblock Unattainability of carnot efficiency in the brownian heat engine.
\newblock {\em Phys. Rev. E}, 62:6021--6025, 2000.

\bibitem{Benenti2011}
G.~Benenti, K.~Saito, and G.~Casati.
\newblock Thermodynamic bounds on efficiency for systems with broken
  time-reversal symmetry.
\newblock {\em Phys. Rev. Lett.}, 106:230602, 2011.

\bibitem{Allahverdyan2013}
A.~E. Allahverdyan, K.~V. Hovhannisyan, A.~V. Melkikh, and S.~G. Gevorkian.
\newblock Carnot cycle at finite power: Attainability of maximal efficiency.
\newblock {\em Phys. Rev. Lett.}, 111:050601, 2013.

\bibitem{Verley2014The}
G.~Verley, M.~Esposito, T.~Willaert, and den Broeck C~Van.
\newblock The unlikely carnot efficiency.
\newblock {\em Nature Communications}, 5:4721, 2014.

\bibitem{Mart2016Brownian}
I.~A. Mart\'{\i}nez, Rold\'{a}n, L~Dinis, D~Petrov, J.~M. Parrondo, and R.~A.
  Rica.
\newblock Brownian carnot engine.
\newblock {\em Nature Physics}, 12(1):67--70, 2016.

\bibitem{Lee2017Carnot}
J.~S. Lee and H.~Park.
\newblock Carnot efficiency is reachable in an irreversible process.
\newblock {\em Scientific Reports}, 7(1):10725, 2017.

\bibitem{Johnson2017Approaching}
C.~V. Johnson.
\newblock An exact model of the power/effciency trade-off while approaching the
  carnot limit.
\newblock {\em arXiv:1703.06119v3}, 2017.

\bibitem{Polettini2017}
M.~Polettini and M.~Esposito.
\newblock Carnot efficiency at divergent power output.
\newblock {\em Europhys. Lett.}, 118(4):40003, 2017.

\bibitem{Pietzonka2018}
P.~Pietzonka and U.~Seifert.
\newblock Universal trade-off between power, efficiency, and constancy in
  steady-state heat engines.
\newblock {\em Phys. Rev. Lett.}, 120:190602, 2018.

\bibitem{Holubec2018}
V.~Holubec and A.~Ryabov.
\newblock Cycling tames power fluctuations near optimum efficiency.
\newblock {\em Phys. Rev. Lett.}, 121:120601, 2018.

\bibitem{Schmiedl2007}
T.~Schmiedl and U.~Seifert.
\newblock Optimal finite-time processes in stochastic thermodynamics.
\newblock {\em Phys. Rev. Lett.}, 98:108301, 2007.

\bibitem{Holubec2014}
V.~Holubec.
\newblock An exactly solvable model of a stochastic heat engine: optimization
  of power, power fluctuations and efficiency.
\newblock {\em J. Stat. Mech.}, 5(5), 2014.

\bibitem{Holubec2015}
V.~Holubec and A.~Ryabov.
\newblock Efficiency at and near maximum power of low-dissipation heat engines.
\newblock {\em Phys. Rev. E}, 92:052125, 2015.

\bibitem{Then2008}
H.~Then and A.~Engel.
\newblock Computing the optimal protocol for finite-time processes in
  stochastic thermodynamics.
\newblock {\em Phys. Rev. E}, 77:041105, 2008.

\bibitem{Horowitz2018}
J.~M. Horowitz and A.~P. Solon.
\newblock Phase transition in protocols minimizing work fluctuations.
\newblock {\em Phys. Rev. Lett.}, 120:180605, 2018.

\bibitem{Schmiedl2008Efficiency}
T.~Schmiedl and U.~Seifert.
\newblock Efficiency of molecular motors at maximum power.
\newblock {\em Europhys. Lett.}, 83(3):30005, 2008.

\bibitem{Zhang20091}
Y.~Zhang.
\newblock The efficiency of molecular motors.
\newblock {\em J. Stat. Phys.}, 134:669--679, 2009.

\end{thebibliography}
\end{document}